\documentclass[prb,amsmath,amssymb,twocolumn,showpacs,floatfix]{revtex4}
\usepackage{graphicx}
\usepackage{bm}

\begin{document}

\title{Curie-like paramagnetism due to incomplete Zhang-Rice singlet formation in La$_{2-x}$Sr$_x$CuO$_4$}

\author{C. V.~Kaiser,$^1$}
\author{W.~Huang,$^1$}
\altaffiliation{Current address: Department of Physics and Astronomy, McMaster University, Hamilton, Ontario L8S 4M1, Canada}
\author{S.~Komiya,$^2$ N. E.~Hussey,$^3$ T.~Adachi,$^4$}
\author{Y.~Tanabe,$^4$}
\altaffiliation{Present address: Advanced Institutes for Materials Research (WPI-AIMR), Tohoku University, Sendai 980-8578, Japan} 
\author{Y.~Koike,$^4$and J. E.~Sonier$^{1,5}$}
\affiliation{$^1$Department of Physics, Simon Fraser University, Burnaby, British Columbia V5A 1S6, Canada \\
$^2$Central Research Institute of Electric Power Industry, Yokosuka, Kanagawa 240-0196, Japan \\
$^3$H.H. Wills Physics Laboratory, University of Bristol, Bristol BS8 1TL, United Kingdom \\
$^4$Department of Applied Physics, Graduate School of Engineering, Tohoku University, Sendai 980-8579, Japan \\
$^5$Canadian Institute for Advanced Research, 180 Dundas Street West, Toronto, Ontario M5G 1Z8, Canada}


\begin{abstract}
In an effort to elucidate the origin of the Curie-like paramagnetism that is generic for heavily-overdoped cuprates, we have performed high 
transverse-field muon spin rotation (TF-$\mu$SR) measurements of La$_{2-x}$Sr$_x$CuO$_4$ single crystals over the Sr content range 
$0.145 \! \leq \! x \! \leq \! 0.33$. We show that the $x$-dependence of the previously observed field-induced broadening of the internal 
magnetic field distribution above the superconducting transition temperature $T_c$ reflects the presence of two distinct contributions.
One of these becomes less pronounced with increasing $x$ and is attributed to diminishing antiferromagnetic correlations. The other
grows with increasing $x$, but decreases above $x \! \sim \! 0.30$, and is associated with the Curie-like term in the bulk magnetic 
susceptibility $\chi$. In contrast to the Curie-like term, however, this second contribution to the TF-$\mu$SR line width extends back into 
the underdoped regime. Our findings imply a coexistence of antiferromagnetically correlated and paramagnetic moments, with the
latter becoming dominant beyond $x \! \sim \! 0.185$. This suggests that the doped holes do not neutralize all Cu spins 
via the formation of Zhang-Rice singlets. Moreover, the paramagnetic component of the
TF-$\mu$SR line width is explained by holes progressively entering the Cu $3d_{x^2-y^2}$ orbital with doping.  
\end{abstract}


\maketitle

\section{Introduction}

In the heavily-overdoped regime of hole-doped cuprates, $T_c$ decreases with
increasing hole concentration $p$, the normal-state pseudogap vanishes, and a Curie-like temperature 
dependence of the normal-state bulk\cite{Takagi:89,Torrance:89,Oda:91,Kubo:91,Allgeier:93,Nakano:94,Levin:96,Schegolev:96,Bras:02,Wakimoto:05} 
and local\cite{Zheng:93,Noc:94,Bellot:97,Williams:98,Chen:08} spin susceptibilities appears. 
In La$_{2-x}$Sr$_x$CuO$_4$ (LSCO) where $p \! = \! x$, Oda {\it et al.}\cite{Oda:91} reported a Curie-like term appearing in
the bulk magnetic susceptibility $\chi$ at $x \! \sim \! 0.18$, becoming more pronounced with increasing $x$, but weakening beyond $x \! \sim \! 0.30$.
They attributed the Curie-like contribution to localized moments that break Cooper pairs and drive down $T_c$. 
However, the origin of such localized moments has never been established, and more recently this has prompted alternative
explanations for the Curie-like behavior of $\chi$.\cite{Kopp:07} 
While often attributed to magnetic impurities or oxygen-disorder induced defects, the Curie-like susceptibility appears to be a universal property of 
heavily-overdoped cuprates, which has persisted through significant improvements in sample quality. 
In fact Takagi {\it et al.}\cite{Takagi:89} demonstrated early on that oxygen vacancies are not directly responsible
for the Curie-like behavior. Nuclear magnetic resonance (NMR) studies of overdoped Tl$_{1-x}$Pb$_x$Sr$_2$CaCu$_2$O$_7$ 
(Ref.~\onlinecite{Bellot:97}) and Bi$_2$Sr$_2$CaCu$_2$O$_{8+\delta}$ (Ref.~\onlinecite{Chen:08}) indicate that the source is 
paramagnetic moments localized in the CuO$_2$ planes. Wakimoto {\it et al.}\cite{Wakimoto:05} have proposed 
that beyond $x \! = \! 0.22$ where a Curie-like term appears in their measurements of $\chi$, that 1/4 of the overdoped Sr ions 
may create the paramagnetic moments either via direct substitution of Sr on the Cu sites, or as a more plausible scenario, the holes 
they add enter the Cu 3$d_{x^2-y^2}$ orbitals. In either case, some fraction of the localized Cu magnetic moments 
are neutralized, breaking up the local AF order and liberating free Cu spins. 

There is some recent evidence for holes predominantly entering the Cu 3$d$ orbitals from Compton scattering measurements
on a LSCO $x \! = \! 0.30$ sample.\cite{Sakurai:11} This is a drastic departure from the widely accepted view that the
doped holes mainly enter the O 2$p$ band, creating Zhang-Rice singlets comprised of a Cu $3d_{x^2-y^2}$ hole state
hybridized with a coherent superposition of O $2p_{x, y}$ orbitals from the four surrounding O atoms.\cite{Zhang:88} 
The formation of Zhang-Rice singlets justifies the use of an effective single-band Hubbard or $t$-$J$ model 
for a satisfactory description of the low-energy physics of the cuprates, rather than a three-band model containing
both the in-plane Cu $3d_{x^2-y^2}$ and O $2p_{x, y}$ orbitals.   
Based on x-ray absorption spectroscopy measurements of LSCO and 
Tl$_2$Ba$_2$CuO$_{6+\delta}$, Peets {\it et al.}\cite{Peets:09} have proposed yet another scenario, whereby the doped 
holes in the O 2p$_{x,y}$ orbitals cease to hybridize with the Cu 3d$_{x^2 - y^2}$ states beyond $p \! \sim \! 0.20$. 
This would result in a loss of the oxygen-mediated in-plane AF superexchange interaction between the localized Cu spins. 
However, this is incompatible with neutron scattering measurements on LSCO that show some remnants of AF correlations 
persisting to at least $x \! = \! 0.25$.\cite{Wakimoto:07} Hence it is likely that some degree of hybridization
persists in LSCO beyond $x \! \sim \! 0.20$.  

Here we report high TF-$\mu$SR measurements on LSCO single crystals. Previous measurements of this
kind revealed a $T$-dependent heterogeneous field response extending far above 
$T_c$.\cite{Savici:05,MacDougall:06,Ishida:07,Sonier:08,MacDougall:10} Through a systematic study of the 
width of the internal magnetic field distribution above $T_c$ as a function of Sr content $x$, MacDougall 
{\it et al.}\cite{MacDougall:10} concluded that the most likely source of the heterogeneous line 
broadening are regions of staggered magnetization about the dopant Sr ions. 
In the present study we show that the $T$-dependent line broadening is associated with the Curie-like term
in the bulk magnetic susceptibility. Consequently, a universal explanation is necessary to explain a
similar $p$-dependent Curie term in $\chi$ for Tl$_2$Ba$_2$CuO$_{6 + \delta}$,\cite{Kubo:91} which is hole doped by increasing the
oxygen content $\delta$. In contrast to bulk magnetic susceptibility studies of LSCO, 
we make no assumptions about the $T$-dependent AF component to isolate the contribution of the localized
paramagnetic moments. As a consequence, we find that the localized moments responsible for the Curie-like
behavior of $\chi$ in heavily-overdoped samples of LSCO also exist at lower hole doping --- gradually becoming more
pronounced with increasing $x$. We attribute the lack of a Curie-like term in $\chi$ below $x \! \sim \! 0.18$ to
limitations of the assumed scaling function introduced by Johnston \cite{Johnston:89} to account for the 
AF correlations, and associate the paramagnetic moments with doped holes entering the Cu $3d_{x^2-y^2}$
orbital.
 
\section{Experimental Details}

\subsection{Samples}

The LSCO single crystals studied here were cut from travelling-solvent floating zone growth rods and subsequently annealed
at elevated temperature and under oxygen partial pressure to minimize oxygen defects. The crystals were prepared 
by three different groups as follows: (i) Single LSCO crystals of Sr content $x \! = \! 0.145$, 0.15, 
0.166, 0.176, 0.19, 0.196, 0.216, and 0.24 with 
$T_c$ = 37.3, 37.6, 37.3, 37.1, 31.4, 30.5, 28, and 17~K, respectively, were grown
by S.~Komiya. These crystals were annealed at 800 to 900$^\circ$C under low oxygen partial pressure (0.2 to 3.5~atm) for 200 hours.
Measurements were also performed on an as-grown $x \! = \! 0.19$ single crystal prepared by S.~Komiya,
which has a substantially reduced and very broad superconducting transition temperature of $T_c \! = \! 23 \! \pm \! 5$~K.
The $x \! = \! 0.15$, 0.166, 0.216, and 0.24 crystals were studied by zero-field (ZF) $\mu$SR in Refs.~\onlinecite{Sonier:10} and 
\onlinecite{Huang:12}, and TF-$\mu$SR measurements of the $x \! = \! 0.166$ and 0.176 crystals were reported in Refs.~\onlinecite{Sonier:08}
and \onlinecite{Sonier:07}.
(ii) Six LSCO single crystals with $x \! = \! 0.26$ and $T_c \! = \! 12 \! \pm \! 2$~K, and two single crystals with $x \! = \! 0.30$ were
synthesized by T.~Adachi and Y.~Tanabe. The $x \! = \! 0.26$ crystals were oxygen annealed at 1~atm and 900~$^\circ$C for 50
hours, and subsequently slow cooled and kept at 500~$^\circ$C for 50 hours.
From chemical titration, the oxygen defect $\delta$ (defined as La$_{2-x}$Sr$_x$CuO$_{4-\delta}$) is estimated
to be $0.014 \! \pm \! 0.01$. The $x \! = \! 0.30$ crystals were oxygen annealed at 3 atm and 900~$^\circ$C for 100
hours. Although we have not estimated $\delta$ for the $x \! = \! 0.30$ crystals, empirically
we believe it is similar to the $x \! = \! 0.26$ sample. We note that bulk magnetic susceptibility measurements
reveal superconducting diamagnetism in a tiny volume fraction of the $x \! = \! 0.30$ crystals below $T_c \! \sim \! 25$~K. 
(iii) Two LSCO single crystals with $x \! = \! 0.33$ were supplied by N.E.~Hussey. One of these was a 380~mg as-grown crystal.
Bulk magnetic susceptibility measurements of this crystal down to 2~K show no signs of superconductivity. The other 33.7~mg
single crystal was annealed for two weeks under an extreme oxygen partial pressure of 400~atm. Resistivity measurements show
that this crystal exhibits no trace of superconductivity down to 0.1~K. Low-temperature ZF-$\mu$SR measurements have revealed
a spin-freezing transition of unknown origin at 0.9~K.\cite{Sonier:10}

\subsection{Muon spin rotation}

The TF-$\mu$SR experiments were carried out at TRIUMF in Vancouver, Canada, using the HiTime spectrometer, which
consists of an ultra low-background sample holder contained inside a helium gas flow cryostat and
a horizontal warm bore 7~T superconducting magnet. Measurements at temperatures above $T \! \sim \! 2$~K were performed with an external 
magnetic field applied perpendicular to the direction of the initial muon spin polarization {\bf P}(0), and 
parallel to the $c$-axis ($\parallel \! {\bf c}$) of the LSCO crystals. 
The TF-$\mu$SR spectra for each sample are well described by the sum of a small time-independent background component 
from muons that miss the sample and evade the background suppression system, and the following power-exponential 
depolarization function:
\begin{equation}
P_i(t) = e^{-(\Lambda t)^\beta} \cos(\gamma_\mu B t + \phi_i) \, , 
\label{eq:pol}
\end{equation}
where $B$ is the magnitude of the average local magnetic field sensed by muons stopping in the sample, and $\phi_i$ is the 
phase angle between {\bf P}(0) and the axis of the $i^{\rm th}$ positron detector ($i \! = \! 1$, 2, 3 and 4).
The power exponential describes the relaxation of the TF-$\mu$SR signal, which occurs when there is a distribution of 
internal magnetic field. For static fields, a larger relaxation rate $\Lambda$ signifies a broader field
distribution, often referred to as inhomogeneous line broadening. If the local magnetic field sensed by each
muon also fluctuates during its lifetime, motional narrowing of the line width occurs, such that $\Lambda$ is reduced 
with increasing fluctuation rate. It is generally not possible to distinguish between static and dynamic 
depolarization of the TF-$\mu$SR signal.  

\section{Results}

\subsection{Field-induced broadening}

\begin{figure}
\centering
\includegraphics[width=8.0cm]{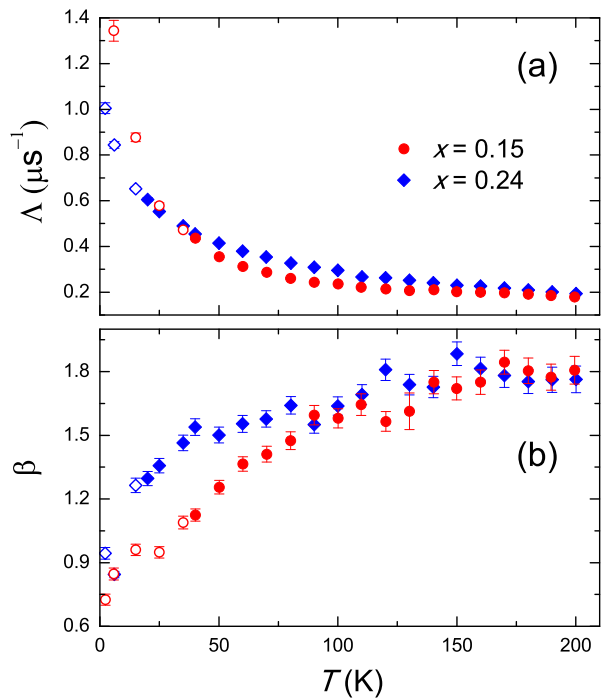}
\caption{(Color online) Temperature dependence of (a) the relaxation rate $\Lambda$, and (b) the power $\beta$
from fits of the TF-$\mu$SR signals of the $x \! = \! 0.15$ and $x \! = \! 0.24$ samples at $H \! = \! 7$~T to Eq.~(\ref{eq:pol}).
The open symbols denote data at temperatures below $T_c$.}
\label{fig:TFx15x24}
\end{figure}

\begin{figure}
\centering
\includegraphics[width=9.0cm]{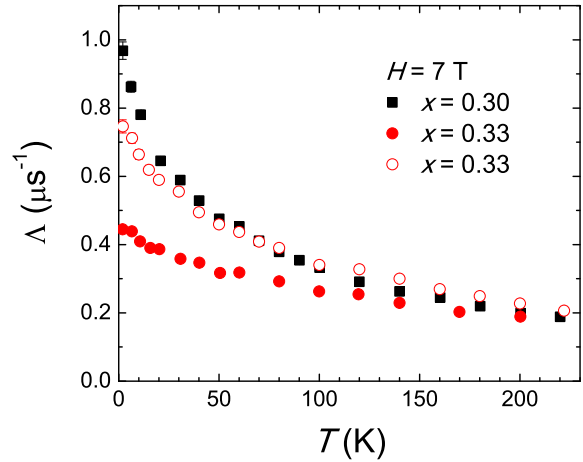}
\caption{(Color online) Comparison of the temperature dependence of the relaxation rate $\Lambda$ at $H \! = \! 7$~T
for the as-grown $x \! = \! 0.33$ single crystal (open red circles) and the $x \! = \! 0.33$ single crystal post annealed under
high-oxygen partial pressure (solid red circles). Also shown is data for the lower pressure annealed $x \! = \! 0.30$ sample.}
\label{fig:x33compare}
\end{figure}

Figure~\ref{fig:TFx15x24} shows representative results of fits of the $x \! = \! 0.15$ and $x \! = \! 0.24$ TF-$\mu$SR signals
at $H \! = \! 7$~T to Eq.~(\ref{eq:pol}). The relaxation rate $\Lambda$ for all samples exhibits a Curie-like temperature dependence 
above $T_c$ indicative of paramagnetic moments. The saturated value $\beta \! \sim \! 1.8$ above $T \! \sim \! 150$~K 
[see Fig.~\ref{fig:TFx15x24}(b)] indicates that the depolarization of the TF-$\mu$SR signal at these temperatures is dominated
by the dipolar fields of the nuclear moments, whereas the fluctuation rate of the electronic moments becomes too fast to
cause significant dynamic depolarization.  
Below $T_c$ there is an additional depolarization associated with the static inhomogeneous line broadening by
vortices, which may or may not form a highly-ordered lattice. The vortex line broadening is proportional to 
$\lambda_{ab}^{-2}$, where $\lambda_{ab}$ is the in-plane magnetic penetration depth.\cite{Sonier:00} 
The effect of the high-oxygen pressure post annealing on the temperature dependence of $\Lambda$ at $x \! = \! 0.33$ is shown 
in Fig.~\ref{fig:x33compare}. The difference between $\Lambda$ in the as-grown and annealed crystals grows significantly with decreasing temperature.
The higher values of $\Lambda$ for the as-grown crystal is ascribed to a spatial distribution of
hole doping caused by oxygen inhomogeneity, and in particular, electronic moments from more underdoped regions of the sample.                 

\begin{figure}
\centering
\includegraphics[width=8.0cm]{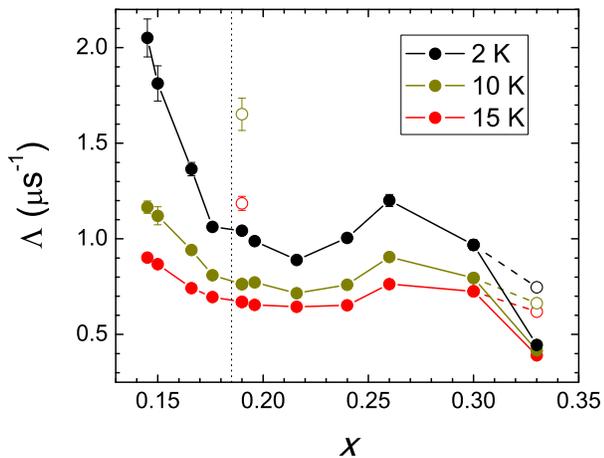}
\caption{(Color online) Dependence of $\Lambda$ on Sr concentration $x$ at $H \! = \! 7$~T for temperatures below $T_c$. The open circles 
are data for the as-grown $x \! = \! 0.19$ and $x \! = \! 0.33$ single crystals. The dashed vertical line indicates $x \! = \! 0.185$.}
\label{fig:BelowTcRelax}
\end{figure}

Figure~\ref{fig:BelowTcRelax} shows the dependence of $\Lambda$ on Sr concentration $x$ at low temperatures.
Below $x \! \sim \! 0.185$ the dependence of $\Lambda$ on $x$ is opposite to the hole-doping dependence of the zero-temperature
extrapolated quantity $\lambda_{ab}^{-2}(0)$.\cite{Panagopoulos:03,Lemberger:11}
This indicates that the width of the internal magnetic field distribution below $x \! \sim \! 0.185$ is not
dominated by static vortices. While the decrease of $\Lambda$ with increasing $x$ could be caused by motional
narrowing of the line width due to rapid fluctuations of the vortices on the $\mu$SR time scale, 
$\Lambda$ (and hence the line width) below $T_c$ increases with increasing magnetic field,\cite{Sonier:08,Sonier:10b} which is contrary
to either a pure vortex-solid or vortex-liquid state. Neutron scattering measurements on LSCO at $H \! = \! 0$ show that remnant
AF spin correlations of the parent insulator La$_2$CuO$_4$ vanish between $x = 0.25$ and $x \! = \! 0.30$.\cite{Wakimoto:07} 
At low temperatures the $H \! = \! 7$~T applied field appears to sufficiently stabilize the AF spin fluctutaions, 
such that the temporal fluctuations of the local magnetic field sensed by the muon do
not average to zero. Here we note that while elastic neutron diffraction experiments indicate that a field of $H \! \ge \! 13$~T is
necessary to induce coexisting static AF order and superconductivity in $x \! \ge \! 0.145$ samples,\cite{Chang:08}
non-uniform static magnetism is observed in the vortex state by TF-$\mu$SR at much lower fields.\cite{Sonier:07}
The latter is explained by interlayer coupling that enables fluctuating magnetism to be stabilized by weakly
interacting vortices.\cite{Kivelson:02} With this said, the decrease of $\Lambda$ with increasing $x$ in Fig.~\ref{fig:BelowTcRelax}
is explained by an increase in the Cu spin fluctuation rate, which reduces the dynamic depolarization of the
TF-$\mu$SR signal.

Just above $x \! \sim \! 0.185$, the near saturation of $\Lambda$ in Fig.~\ref{fig:BelowTcRelax} indicates that the influence
of the AF spin fluctuations on the TF-$\mu$SR line width becomes negligible. Since measurements of 
$\lambda_{ab}^{-2}(0)$ in $0.07 \! \leq \! x \! \leq \! 0.24$ LSCO powders by Panagopoulos 
{\it et al.}\cite{Panagopoulos:03} show a saturation of $\lambda^{-2}(0)$ above $x \! = \! 0.19$, it appears that
the TF-$\mu$SR linewidth immediately above $x \! \sim \! 0.185$ is dominated by the static inhomogeneous field distribution of 
the vortex-solid phase. However, recent measurements of $0.06 \! \leq \! x \! \leq \! 0.30$ LSCO films by Lemberger {\it et al.},\cite{Lemberger:11}
show $\lambda_{ab}^{-2}(0)$ decreasing with increasing $x$ above $x \! = \! 0.19$, and vanishing at $x \! = \! 0.27$.
The different $x$-dependences of $\lambda_{ab}^{-2}(0)$ reported in Refs.~\onlinecite{Panagopoulos:03} and \onlinecite{Lemberger:11} 
may be due to varying degrees of phase separation into superconducting and nonsuperconducting metallic regions
--- typical of heavily-overdoped LSCO,\cite{Tanabe:05,Wang:07,Adachi:09} and other cuprates.\cite{Wen:02}
While the origin of this discrepancy is uncertain, the observed increase of $\Lambda$ beyond $x \! \sim \! 0.216$ in Fig.~\ref{fig:BelowTcRelax} divulges  
another source of line width broadening superimposed on the effect of the vortices. Above the usual 
superconducting-to-nonsuperconducting phase transition at $x \! = \! 0.27$, the vortex contribution vanishes and the value of $\Lambda$ 
shown in Fig.~\ref{fig:BelowTcRelax} is reduced. Note that while there is a trace amount of superconductivity in the $x \! = \! 0.30$ sample, 
the corresponding regions are likely not superconducting at $H \! = \! 7$~T.       

\begin{figure}
\centering
\includegraphics[width=9.0cm]{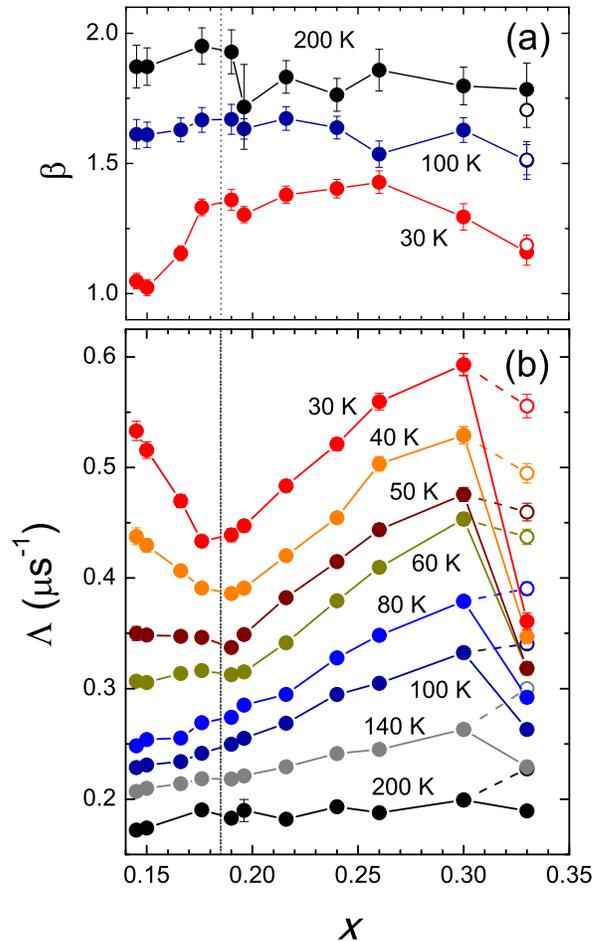}
\caption{(Color online) (a) Dependence of $\beta$ and (b) $\Lambda$ on Sr concentration $x$ at $H \! = \! 7$~T and temperatures $T \! \geq \! 30$~K. 
The open circles are data for the as-grown $x \! = \! 0.33$ single crystal. The dashed vertical line indicates $x \! = \! 0.185$.}
\label{fig:AboveTcRelax}
\end{figure}

\begin{figure}
\centering
\includegraphics[width=9.0cm]{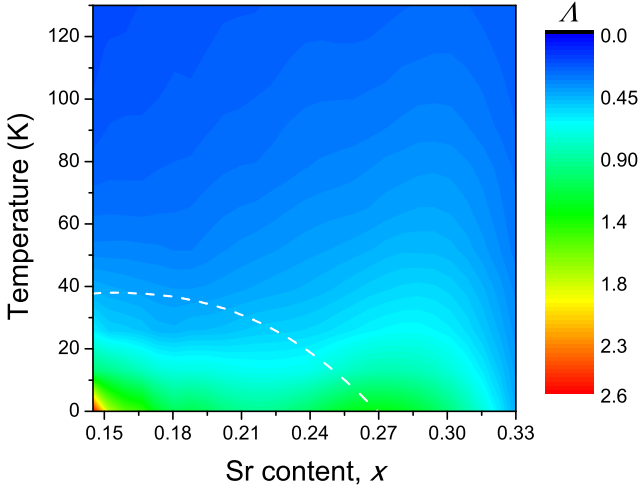}
\caption{(Color online) Contour plot of the variation of the relaxation rate $\Lambda$ at
$H \! = \! 7$~T with temperature and Sr content in the annealed LSCO single crystals. The
phase diagram is generated by interpolating the $\Lambda$ versus $T$ data for each sample.
The dashed white curve approximately indicates $T_c$ at $H \! = \! 0$~T.}
\label{fig:ContourRelax}
\end{figure}

Above $T_c$ where the vortex contribution is absent, the component of $\Lambda$ that increases with increasing $x$ is more 
obvious and is seen to intensify up to $x \! = \! 0.30$ [see Fig.~\ref{fig:AboveTcRelax}(b)]. At temperatures $T \! > \! 60$~K it becomes 
apparent that this component of $\Lambda$ extends back to the lowest doping at $x \! = \! 0.145$. Hence it is clear
that the source of this behavior is not restricted to heavily overdoped phase-separated samples, and does not onset at $x \! \sim \! 0.185$.
Instead $x \! \sim \! 0.185$ appears to mark a low-temperature crossover from a hole-doping range where the dominant
electron spin contribution to $\Lambda$ comes from remnant AF correlations of the parent compound, 
to a higher hole-doping range dominated by a different kind and/or arrangement of magnetic moments. 
The slight increase of $\Lambda$ with increasing $x$ at $T \! = \! 200$~K in Fig.~\ref{fig:AboveTcRelax}(b) 
implies that the line broadening at this temperature is dominated by the nuclear dipole moments.
This conclusion is also reached from inspecting the corresponding values of $\beta$ in Fig.~\ref{fig:AboveTcRelax}(a).
A complete summary of our measurements of $\Lambda$ as function of $T$ and $x$ is shown in Fig.~\ref{fig:ContourRelax}.

\subsection{Magnetic susceptibility and muon Knight shift}

\begin{figure}
\centering
\includegraphics[width=9.0cm]{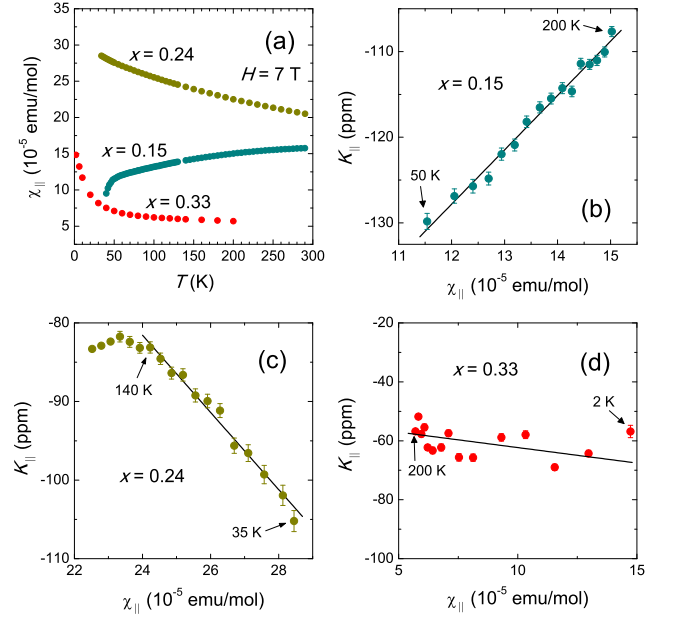}
\caption{(Color online) (a) Temperature dependence of the bulk molar susceptibility above $T_c$ for a magnetic field 
$H \! = \! 7$~T applied parallel to the $c$-axis. (b) to (d) Plots of the normal-state 
muon Knight shift versus the bulk molar susceptibility, with temperature as an implicit parameter.}
\label{fig:Knight}
\end{figure}

Figure~\ref{fig:Knight}(a) shows the temperature dependence of the bulk magnetic susceptibility of the 
annealed $x \! = \! 0.15$, $x \! = \! 0.24$ and $x \! = \! 0.33$ single cystals for ${\bf H} \! \parallel \! {\bf c}$. 
For $x \! > \! 0.08$ the magnetic susceptibility of LSCO in the normal state has previously been described by the following equation:
\begin{equation}
\chi(x, T) = \chi^{\rm 2D}(x, T) + \chi_0(x) + C(x)/T \, .
\label{eq:susceptibility}
\end{equation} 
The contribution $\chi^{\rm 2D}(x, T)$ follows a 
simple universal scaling relation,\cite{Johnston:89} attributed to residual two-dimensional (2D) AF correlations in
the CuO$_2$ planes. In particular, $\chi^{\rm 2D}(x,T)/\chi^{\rm 2D}_{\rm max} \! = \! F(T/T_{\rm max})$, where 
$\chi^{\rm 2D}_{\rm max}$ and $T_{\rm max}$ are $x$-dependent values corresponding to the maximum of $\chi^{\rm 2D}(x, T)$,
and $F$ is a universal curve.     
The $T$-independent term $\chi_0(x)$ in Eq.~(\ref{eq:susceptibility}) is the sum of the atomic core diamagnetism $\chi_{\rm core}$,
the anisotropic Van Vleck paramagnetism $\chi_{\rm VV}$ and the Pauli paramagnetism $\chi_{\rm Pauli}$. As discussed later, 
the $x$ dependence of $\chi_0$ is caused by a variation of $\chi_{\rm Pauli}$. The third term in Eq.~(\ref{eq:susceptibility})
has a Curie-like $T$-dependence that appears above $x \! \sim \! 0.18$.\cite{Oda:91,Nakano:94,Wakimoto:05}      
The Curie constant $C(x)$ increases with increasing $x$ above $x \! \sim \! 0.18$, signifying the growing presence of localized 
paramagnetic moments, but decreases beyond $x \! \sim \! 0.30$.    

The local spin susceptibility causes a muon Knight shift, which is defined as the fractional difference between the average 
magnetic field $B$ at the muon site and the applied field $H$. Correcting for macroscopic contributions to $B$ that are present 
in the external field, the Knight shift originating from microscopic contributions in the sample is
\begin{equation}
K = \frac{B - H}{H} - K_{\rm dem,L} \, ,
\label{eq:Knight}
\end{equation}
where $K_{\rm dem,L} \! = \! 4 \pi (1/3-N) \rho_{\rm mol} \chi_\parallel$ is a correction 
for demagnetization and Lorentz fields, $N$ is the demagnetization factor for the sample, 
$\rho_{\rm mol}$ is the molar density of the sample in units of mol/cm$^3$, and $\chi_\parallel$ 
is the bulk molar susceptibility displayed in Fig.~\ref{fig:Knight}(a). The applied field induces spin polarization of both the conduction 
electrons and localized electronic moments, such that
\begin{equation}
K =  K_0 + \frac{1}{H^2} {\bf H} \cdot \stackrel{\leftrightarrow}{A}_{\rm eff} \cdot \stackrel{\leftrightarrow}{\chi} \cdot {\bf H} \, .
\label{eq:KnightHyperfine}
\end{equation}   
The first term is a $T$-independent contribution from the Pauli spin paramagnetism
of the conduction electrons that screen the $\mu^+$ charge. The second term is $T$-dependent and arises from the polarization
of localized electronic moments. It is expressed in terms of an effective hyperfine coupling tensor 
$\stackrel{\leftrightarrow}{A}_{\rm eff}$ and the field-induced localized moment $\stackrel{\leftrightarrow}{\chi} \cdot {\bf H}$.
The effective hyperfine coupling $\stackrel{\leftrightarrow}{A}_{\rm eff}$ is the sum of dipolar
$\stackrel{\leftrightarrow}{A}_{\rm dip}$ and hyperfine contact coupling $\stackrel{\leftrightarrow}{A}_{\rm c}$ contributions.
The elements of the dipolar coupling tensor $\stackrel{\leftrightarrow}{A}_{\rm dip}$ depend 
on the $\mu^+$ site and the crystallographic structure, and are given by
\begin{equation}
A_{\rm}^{ij} = \sum_{\rm local} \frac{1}{r^3} \left( 3 \frac{r_i r_j}{r^2} - \delta_{ij} \right) \, ,
\label{dipolar}
\end{equation}
where the sum extends over all localized moments inside a Lorentz sphere, and $r$ is the magnitude
of the vector ${\bf r} \! = \! (x, y, z)$ connecting the muon site to the localized moment site
in the crystal lattice.  
The hyperfine contact coupling, which is often considered as isotropic and $T$-independent, results from further spin 
polarization of the conduction electrons by induced localized moments via the Rudermann-Kittel-Kasuya-Yoshida (RKKY) interaction.
In cuprates, an RKKY interaction may occur between localized Cu spins
and doped itinerant holes. Calculations show that the RKKY interaction is enhanced with increasing hole concentration, and may 
dominate over the antiferromagnetic superexchange coupling in the overdoped regime.\cite{Si:92,Nunez:95,Kolley:98}  

Figures~\ref{fig:Knight}(b) to \ref{fig:Knight}(d) show the normal-state Knight shift (for ${\bf H} \parallel \! {\bf c}$) 
of the $x \! = \! 0.15$, $x \! = \! 0.24$ and $x \! = \! 0.33$ single crystals, corrected for demagnetization and Lorentz fields, 
and plotted versus $\chi_\parallel$. For a wide range of temperature above $T_c$, $K_\parallel$ exhibits linear scaling 
with $\chi_\parallel$, indicating an equivalency of the bulk and local magnetic spin susceptibilities. The saturation of $K_{\parallel}$ at 
high temperatures for the $x \! = \! 0.24$ sample likely signifies the resolution limit of the measured 
TF-$\mu$SR frequency shift. Likewise, the Knight shift
at $x \! = \! 0.33$ appears to be too small to measure accurately with the spectrometer used in our experiments. 

The most notable feature of the $K_\parallel$ versus $\chi_\parallel$ plots is the change in the slope from positive
at $x \! = \! 0.15$ to negative at $x \! = \! 0.24$. A positive slope was previously reported for $x \! = \! 0.07$.\cite{Ishida:07} 
The slope is equal to the effective hyperfine coupling tensor element 
$A_{\rm eff}^{\parallel}$ for ${\bf H} \! \parallel \! {\bf c}$. Unfortunately, the present measurements for a single direction of 
applied magnetic field are insufficient for determining the dipolar and contact hyperfine couplings. 
The problem is compounded by the apparent presence of both AF correlated spins and local paramagnetic
moments. Consequently, to explain the changes in slope we resort to a heuristic argument.
Recently it has been determined that the muon stops near the non-axially symmetric site 
(0.120$a$, 0, 0.219$c$).\cite{Huang:12} At this site and with ${\bf H} \! \parallel \! {\bf c}$, the calculated dipolar coupling constant for 
antiferromagnetically ordered Cu spins (canted by 45$^\circ$ or lying in the $a$-$b$ plane) is $A_{\rm dip}^\parallel \! = \! -0.33$~kG/$\mu_B$.
For parallel Cu spins aligned along the applied field direction, $A_{\rm dip}^\parallel \! = \! 0.97$~kG/$\mu_B$.   
Hence the change in the sign of the slope between $x \! = \! 0.15$ and $x \! = \! 0.24$ can be explained by a change in 
the primary alignment of the Cu spins. However, the contact hyperfine coupling constant $A_{\rm c}^\parallel$ is also
expected to vary with $x$, not only because of a change in the effective exchange interaction between the Cu moment
and the conduction electrons, but also as discussed in the next section, because the density of electronic states at the Fermi energy
is $x$-dependent.

\section{Discussion}

We have shown that the normal state of LSCO single crystals from three different sources is characterized by a broad field-induced
distribution of internal magnetic field. The dependence of the TF-$\mu$SR relaxation rate $\Lambda$ on $x$ indicates two 
distinct $T$-dependent contributions. The first of these decreases with increasing $x$ 
and is clearly associated with diminishing remnant AF correlations. While neutron scattering\cite{Wakimoto:07,Lipscombe:07,Wilson:12}
and ZF-$\mu$SR\cite{Risdiana:08} measurements on pure and Zn-doped
LSCO in zero external magnetic field suggest that AF spin fluctuations likely persist out to the termination 
of superconductivity in the overdoped regime, the effect of AF correlations
on the TF-$\mu$SR line width at $H \! = \! 7$~T is apparent only up to $x \! \sim \! 0.185$.
This is partly due to the diminishing effect of the applied magnetic field in stabilizing the increasingly faster AF fluctuations,
but it is also a consequence of a second contribution to the TF-$\mu$SR line width that grows with increasing $x$. We find, however, 
that this second source of line broadening is greatly reduced when superconductivity completely disappears beyond $x \! = \! 0.30$.

\begin{figure}
\centering
\includegraphics[width=9.0cm]{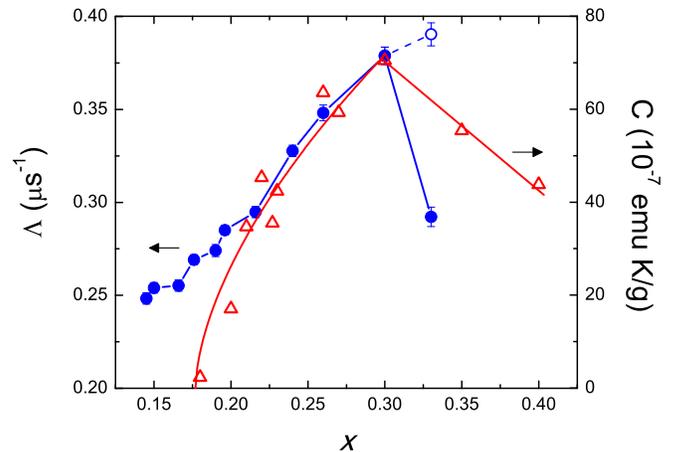}
\caption{(Color online) Comparison of the $x$-dependence of $\Lambda$ at $T \! = \! 80$~K to the $x$-dependence of the Curie constant $C$ from
Ref.~\onlinecite{Oda:91}. The open circle corresponds to $\Lambda$ for the as-grown $x \! = \! 0.33$ single crystal.}
\label{fig:Oda}
\end{figure}

Since the Knight shift $K_\parallel$ exhibits a linear scaling with $\chi_\parallel$ for
a wide temperature range above $T_c$, it is instructive to consider the $\mu$SR results in the context of comprehensive studies of the
normal-state bulk magnetization of LSCO. To begin with, the $T$-independent component of the bulk magnetic susceptibility
$\chi_0$ is strongly dependent on the hole concentration. Measurements on samples in the range $0.10 \! \leq \! x \! \leq \! 0.45$ 
show that $\chi_0$ increases with increased hole doping, plateaus above $x \! \sim \! 0.20$ to 0.23, and decreases above 
$x \! \sim \! 0.30$.\cite{Nakano:94,Nakano:96,Loram:96} While the $T$-independent component of the total static
magnetic susceptibility includes contributions from the core diamagnetism and Van Vleck paramagnetism, the doping 
dependence of $\chi_0$ has been attributed to changes in the Pauli susceptibility $\chi_{\rm Pauli}$.
This is because the $x$ dependences of $\chi_0$ and the normal-state electronic specific heat coefficient $\gamma$ 
are very similar below $x \! \sim \! 0.24$,\cite{Nakano:96} and in a free-electron 
system $\chi_{\rm Pauli}$ and $\gamma$ are each proportional to the density of states $N(E_F)$ at the Fermi energy. 
The $x$ dependences of $\chi_{\rm Pauli}$ and $\gamma$ can be explained by the presence of a van Hove singularity
(VHS) in the electronic density of states caused by the flat part of the conduction bands of LSCO 
being very close to $E_F$. With increasing $x$, the ($\pi$, 0) flatband and VHS move to higher energy and cross $E_F$ at 
$x \! \sim \! 0.20$,\cite{Ino:02,Sahrakorpi:08} where $N(E_F)$,\cite{Ino:98} and hence both $\chi_0$ and $\gamma$ reach their maximum 
values. 

The increase of $\chi_0$ with increasing $x$ up to $x \! \sim \! 0.20$ is accompanied by a 
reduction of the effective moment of the localized Cu spins,\cite{Nakano:94,Nakano:96} due to
the frustration effect of the doped holes in the O 2p$_{x,y}$ orbitals on the AF exchange interaction between Cu spins. 
As mentioned above, the suppression of the AF correlations of the localized Cu spins is the source of
the initial decrease of $\Lambda$ with increasing $x$ in Figs.~\ref{fig:BelowTcRelax}, \ref{fig:AboveTcRelax}(b) and \ref{fig:ContourRelax}.
On the other hand, we attribute the additional $T$-dependent contribution to $\Lambda$ that grows with increasing $x$ to the same source of the
Curie-like constant $C(x)$ in the bulk magnetic susceptibility above $x \! \sim \! 0.18$. To see this is the case, in
Fig.~\ref{fig:Oda} we compare $C(x)$ from Ref.~\onlinecite{Oda:91} to $\Lambda(x)$ at $T \! = \! 80$~K, where the AF contribution is minor.
Note as expected, the interpolated data of Ref.~\onlinecite{Oda:91} falls between the two $\Lambda$ values at $x \! = \! 0.33$, which correspond
to the as-grown and high-oxygen pressure annealed extremes of sample preparation. 
While the $\Lambda(x)$ data points at $x \! = \! 0.33$ in Fig.~\ref{fig:Oda} seem to suggest that the decrease of $C(x)$ 
beyond $x \! = \! 0.30$ is a sole consequence of annealing, the data at temperatures below $T \! = \! 80$~K in Fig.~\ref{fig:AboveTcRelax}(b)
clearly indicates that this is not the case. In fact the $T$-independent component $\chi_0$ of the
bulk magnetic susceptibility of the as-grown crystal at $x \! = \! 0.33$ is larger than that of the annealed crystal,
such that the former resembles an annealed sample at a somewhat lower value of $x$. 

In contrast to $C(x)$, the
second $T$-dependent contribution to $\Lambda(x)$ extends below $x \! = \! 0.18$ into the underdoped regime.
Strictly speaking, the Curie-like term observed in overdoped LSCO in Refs.~\onlinecite{Oda:91} and \onlinecite{Nakano:94} is the 
deviation of the bulk magnetic susceptibility from the assumed $T$-dependent form for $\chi^{\rm 2D}(x, T)$ in Eq.~(\ref{eq:susceptibility}),
which is described by the universal scaling function $F$ introduced by Johnston.\cite{Johnston:89}
Consequently, we ascribe the separation of the data for $C(x)$ and $\Lambda(x)$ below $x \! \sim \! 0.21$ in Fig.~\ref{fig:Oda}
to an inaccuracy in the assumed form for $\chi^{\rm 2D}(x, T)$. In actuality, Eq.~(\ref{eq:susceptibility}) is only valid
if the doped holes in the O $2p_{x,y}$ orbitals are weakly hybridized with the Cu $3d_{x^2-y^2}$ states,\cite{Allgeier:93} 
in which case $\chi_0(x)$ and $\chi^{\rm 2D}(x, T)$ can be treated as distinct terms. According to the     
x-ray absorption spectroscopy measurements by Peets {\it et al.}\cite{Peets:09}, this may be the case
only at $x \! > \! 0.20$, where we find that the $x$-dependence of $\Lambda$ closely follows $C(x)$.     

Based on the following observations we conclude that the paramagnetic moments are the result of 
doped holes that do not stimulate the formation of Zhang-Rice singlets: (i) NMR measurements indicate 
that the source of the Curie-like paramagnetism resides in the CuO$_2$ layers, (ii) the Curie-like behavior is 
observed in cuprates hole-doped by either cation substitution or changes in oxygen content, (iii) we find good agreement 
between samples synthesized by different groups, and (iv) we have demonstrated that the localized
electronic moments responsible also exist in the underdoped regime. 
Recently, Sordi {\it et al.} have shown that doublet formation (total spin $S \! = \! 1/2$) in a one-band Hubbard model
grows upon doping the parent Mott insulator.\cite{Sordi:11} The same is true of the $S \! = \! 1/2$ three-spin polaron
present in three-band models,\cite{Emery:88,Lau:11} where the spin of a hole on the in-plane oxygen is antiparallel
to the two nearest-neighbor copper spins. However, the hole in these and the triplet state (total spin $S \! = \! 1$)
is mobile, with the effective magnetic moment presumably fluctuating too fast to cause dynamic depolarization of the 
TF-$\mu$SR signal.

The paramagnetic moments are more likely caused by a growing number of holes entering the Cu $3d$ orbitals,
as evidenced by recent Compton scattering measurements of LSCO.\cite{Sakurai:11}
Localized spins are created by holes entering the out-of-plane Cu $3d_{z^2-r^2}$ orbital, but
this does not explain the reduction of $C$ (and $\Lambda$) beyond $x \! \sim \! 0.30$.
Doped holes entering the half-filled in-plane Cu $3d_{x^2-y^2}$ orbital neutralize existing Cu spins, 
hence disrupting local AF order and creating free Cu spins. With doping, an increasing number of free Cu spins are 
generated in this way until the disappearance of AF correlations in the overdoped regime. The additional
holes introduced by further doping progressively neutralizes the free Cu $3d_{x^2-y^2}$ spins, in agreement
with the reduction of the Curie-like paramagnetism beyond $x \! \sim \! 0.30$.
This appears to be a universal property of the cuprates, and consequently is a restriction on the applicability of 
one-band models.    
 
We thank G. A. Sawatzky and A.-M.S. Tremblay for informative discussions, and both the technical staff at the TRIUMF 
Centre for Molecular and Materials Science and N. Mangkorntong for technical assistance. This work was supported by 
the Natural Sciences and Engineering Research Council of Canada, and the Canadian Institute for Advanced Research.
N.E. Hussey acknowledges the support of a Royal Society Wolfson Research Merit Award.

\end{document}